\documentclass[reprint,
superscriptaddress,
 aps,
pra,
]{revtex4-1}

\usepackage[utf8]{inputenc}
\usepackage{amsmath}
\usepackage{amsfonts}
\usepackage{braket}
\usepackage{comment}
\usepackage{graphicx}
\usepackage{enumitem,kantlipsum}
\usepackage{physics}
\usepackage{xcolor}
\usepackage{multirow}
\usepackage{adjustbox}
\usepackage{hyperref}

\begin{document}

\preprint{APS/123-QED}
\title{Inhomogeneous Polarization Transformation Reveals PT-Transition in non-Hermitian Optical Beam Shift}

\author{Niladri Modak}
\email{niladri.modak@tuni.fi}
\affiliation{Department of Physical Sciences, Indian Institute of Science Education and Research (IISER) Kolkata, Mohanpur 741246, India.}
\affiliation{Photonics Laboratory, Physics Unit, Tampere University, Tampere 33720, Finland.}
\author{Swain Ashutosh}
\affiliation{Department of Physical Sciences, Indian Institute of Science Education and Research (IISER) Kolkata, Mohanpur 741246, India.}
\affiliation{School of Physics and Astronomy, University of Glasgow, Glasgow, G12 8QQ, United Kingdom.}
\author{Shyamal Guchhait}
\affiliation{Department of Physical Sciences, Indian Institute of Science Education and Research (IISER) Kolkata, Mohanpur 741246, India.}
\author{Sayan Ghosh}
\affiliation{Department of Physical Sciences, Indian Institute of Science Education and Research (IISER) Kolkata, Mohanpur 741246, India.}
\author{Ritwik Dhara}
\affiliation{Department of Physical Sciences, Indian Institute of Science Education and Research (IISER) Kolkata, Mohanpur 741246, India.}
\author{Jeeban Kumar Nayak}
\affiliation{Department of Physical Sciences, Indian Institute of Science Education and Research (IISER) Kolkata, Mohanpur 741246, India.}
\author{Sourin Das}
\affiliation{Department of Physical Sciences, Indian Institute of Science Education and Research (IISER) Kolkata, Mohanpur 741246, India.}
\author{Nirmalya Ghosh}
\email{nghosh@iiserkol.ac.in}
\affiliation{Department of Physical Sciences, Indian Institute of Science Education and Research (IISER) Kolkata, Mohanpur 741246, India.}

\begin{abstract}
\noindent 
Despite its non-Hermitian nature, the transverse optical beam shift exhibits both real eigenvalues and non-orthogonal eigenstates. To explore this unexpected similarity to typical PT (parity-time)-symmetric systems, we first categorize the entire parametric regime of optical beam shifts into Hermitian, PT-unbroken, and PT-broken phases. Besides experimentally unveiling the PT-broken regime, crucially, we illustrate that the observed PT-transition is rooted in the momentum-domain inhomogeneous polarization transformation of the beam. The correspondence with a typical non-Hermitian photonic system is further established. Our work not only resolves a longstanding fundamental issue in the field of optical beam shift but also puts forward the notion of novel non-Hermitian spin-orbit photonics: a new direction to study non-Hermitian physics through the optical beam shifts.

\end{abstract}

\maketitle

\paragraph{Introduction}
Optical beam shifts from dielectric interfaces have played an 
important role in 
enriching our understanding of various light-matter interactions \cite{born2013principles}, simulating and interpreting analogous physical phenomena \cite{modak2023longitudinal,asano2016anomalous}, and also producing optical devices for metrology and sensing \cite{ling2017recent}. There are different platforms that exhibit such beam shifts starting from natural light-matter interactions  \cite{bliokh2013goos}, e.g., partial reflection (PR) \cite{bliokh2013goos}, total internal reflection (TIR) \cite{bliokh2013goos}, transmission from an interface \cite{bliokh2013goos}, or through a tilted anisotropic medium \cite{bliokh2016spin,modak2022tunable} to exotic metastructures \cite{kong2019goos}. The longitudinal shift is called Goos-H\"{a}nchen (GH) shift \cite{bliokh2013goos} and originates from the dispersion of dynamical parameters, e.g., Fresnel coefficients \cite{bliokh2013goos} or anisotropic transmission coefficients \cite{bliokh2016spin}. The transverse shift is called Imbert-Fedorov (IF) shift or spin-Hall shift \cite{bliokh2013goos},\cite{hosten2008observation}. It is associated with more intriguing physical phenomena involving the evolution of the geometric phase \cite{bliokh2013goos} and the spin-orbit interaction of light \cite{hosten2008observation}.

\par
These optical beam shifts are often described by shift matrices \cite{toppel2013goos,gotte2014eigenpolarizations}, also known as the Artman operators \cite{toppel2013goos}, whose eigenvalue corresponds to the magnitude of the beam shifts \cite{gotte2014eigenpolarizations}. 
Although in general, the shift matrices can be non-Hermitian \cite{toppel2013goos,gotte2014eigenpolarizations}, depending on the system parameters it shows some peculiarities. For example, although for TIR, the shift matrix is Hermitian, as one goes from TIR to PR, interestingly, the IF shift matrix changes its Hermiticity \cite{bliokh2013goos,toppel2013goos}. Moreover, for PR, even if the shift matrix remains non-Hermitian, the eigenvalues can still be real, identical to a PT (parity-time) symmetric non-Hermitian system \cite{gotte2014eigenpolarizations}. On the other hand, the non-Hermitian GH shift matrix does not exhibit such a transition (see Supporting Information (SI) Sec. S.1). Although the optical beam shifts have been studied extensively for the last two decades \cite{hosten2008observation,jayaswal2014observation,goswami2014simultaneous,modak2023longitudinal,gotte2014eigenpolarizations,bliokh2013goos,bliokh2016spin,toppel2013goos}, such peculiar non-Hermitian nature and its origin remain largely unexplored. Moreover, the optical beam shifts rooted in the straightforward diffractive corrections of the beam \cite{bliokh2013goos} exhibiting such strong analogies with intricate PT-symmetric non-Hermitian systems also demands a deeper understanding of its origin. In this work, we address this issue by taking an example of IF shift from reflection of a fundamental Gaussian beam. All the explanations and discussions can be trivially extrapolated for all other polarization-dependent beam shifts (e.g., GH shift) in different light-matter interactions due to their common origin in the momentum or position-domain polarization modulation of the beam \cite{bliokh2013goos,toppel2013goos,kong2019goos}.
\par We first segregate the whole parametric regime  into Hermitian, PT-unbroken, and PT-broken phases of IF shifts. We then experimentally detect the momentum-domain eigenshifts in the PT-broken regime, which was not explored earlier. Next and more importantly, we find the origin of the discussed non-Hermiticity in the momentum-domain polarization evolution of the beam and demonstrate that the observed PT-transition \cite{bender1998real,el2018non,krasnok2021parity} stems from the momentum-domain inhomogeneous polarization \cite{lu1994homogeneous} modulation. Moreover, our calculation reveals that such a transition is not limited to the IF shift operator; instead, it occurs for a similar inhomogeneous polarization element for all possible polarization anisotropy effects.
\par Note that the PT-symmetric system has recently found great interest in various areas ranging from condensed matter physics \cite{kreibich2014realizing}, optics \cite{guo2009observation}, and photonics \cite{doppler2016dynamically}
to material science and engineering \cite{lupu2013switching}, especially in the context of non-Hermitian systems. Earlier, losses in physical systems were perceived as undesirable, and measures were taken to avoid them. This is where the alliance between PT-symmetry and non-Hermitian systems offers a great advantage in using or, more so, tuning the loss for various applications \cite{lin2011unidirectional,hodaei2014parity,fleury2015invisible}. Here, we also develop a correspondence with a typical non-Hermitian photonic system, where equal and opposite polarities of pristine spin-Hall shift mimic the two coupled modes with linear polarization-dependent losses representing the coupling. This way, we found that all the signatures of a typical PT-symmetric system can be thoroughly adapted to the optical beam shifts platform. Therefore, our study not only tackles a fundamental issue in the field of optical beam shift but also enables this relatively simple system of a partial reflection of a Gaussian beam to be an effective prototype for investigating typical non-Hermitian systems. 
\par

\begin{figure}[hb!]
\centering
\includegraphics[width=0.8\linewidth]{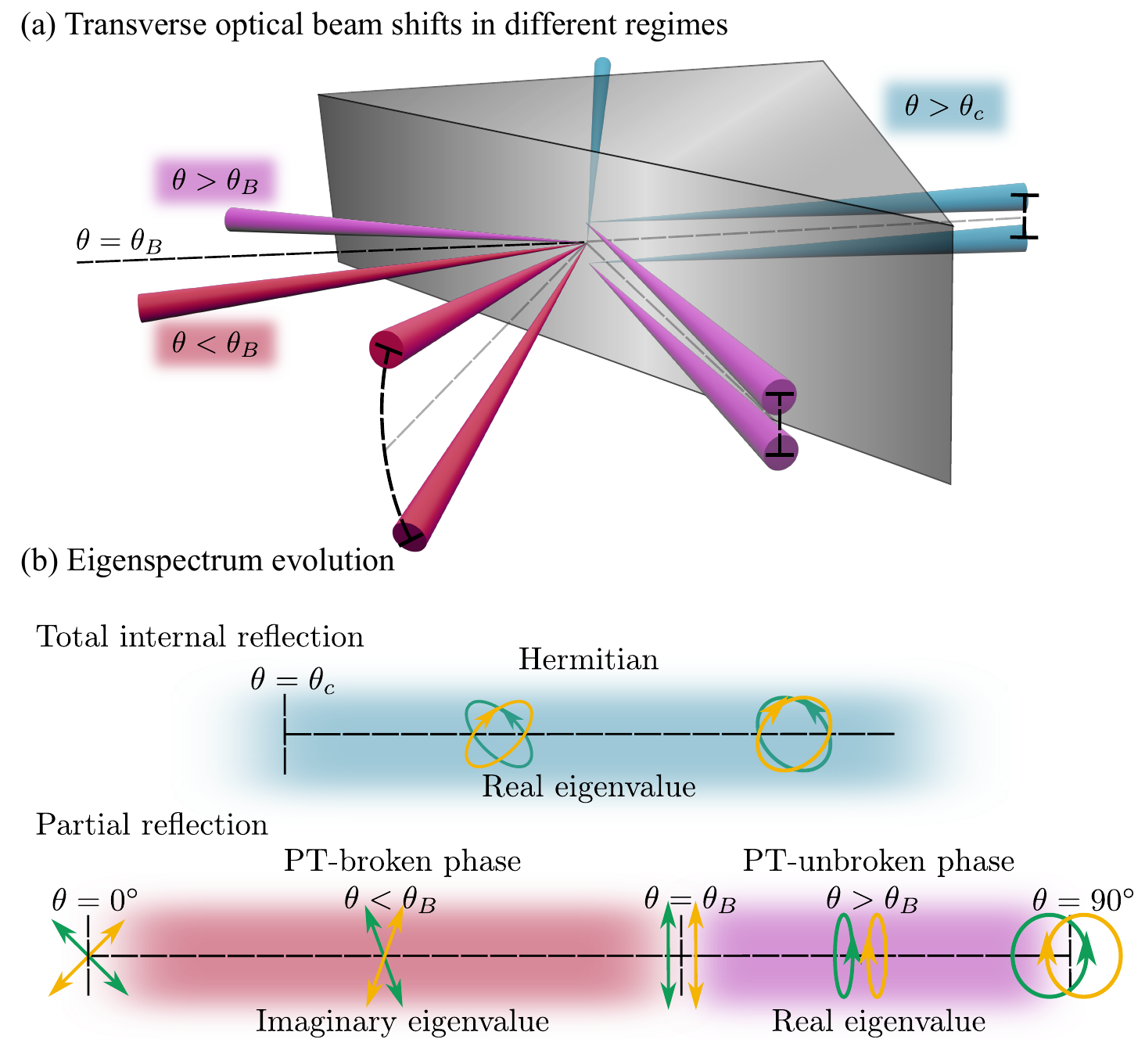}
\caption{\textit{Schematic illustration of the eigenvalues and eigenstates of transverse optical beam shift in different experimental regimes.} For TIR, i.e., $\theta> \theta_c$ (blue beams in (a)), the shift is spatial owing to the real eigenvalues of the corresponding Hermitian shift matrix (blue-shaded area in (b)). The corresponding eigenstates are elliptic and orthogonal (green and yellow ellipses). For PR, the corresponding non-Hermitian shift matrix has real eigenvalues for $\theta>\theta_B$ and imaginary eigenvalues for $\theta<\theta_B$, manifested as spatial (magenta beams in (a)) and momentum-domain shifts (red beams in (a)) respectively. For $\theta>\theta_B$, the non-orthogonal eigenstates are elliptical and become orthogonal circular at $\theta\rightarrow 90^{\circ}$ (magenta-shaded area in (b)). For $\theta<\theta_B$, the non-orthogonal eigenstates are linear and become orthogonal $\pm45^{\circ}$ when $\theta\rightarrow0^{\circ}$ (magenta-shaded area in (b)). At $\theta\rightarrow\theta_B$, the eigenstates become collinear. The transition in the eigenspectrum around $\theta=\theta_B$ resembles a typical PT-transition, i.e., transition of a non-Hermitian system from PT-unbroken to PT-broken phase.}
\label{fig1}
\end{figure}


\paragraph{Hermiticity of IF shift in different parametric regimes}
We start with the well-known shift matrix of the IF shift in the reflection of a light beam with wavenumber $k$ \cite{toppel2013goos,gotte2014eigenpolarizations}.
\begin{equation}
    \hat{A}_y=i\frac{\cot{\theta}}{k}\begin{pmatrix} 0 & (1+r_p/r_s)\\ -(1+r_s/r_p) & 0
\end{pmatrix}
\end{equation}
Here $r_p$ and $r_s$ are the Fresnel reflection coefficients \cite{gupta2015wave} for longitudinal ($\hat{x}$) and transverse ($\hat{y}$) polarized incident light respectively, $\theta$ is the angle of incidence. The matrix $\hat{A}_y$ is, in general, non-Hermitian
with eigenvalues $\lambda_{\pm}$ and eigenstates (right) $|\pm\rangle$ \cite{gotte2014eigenpolarizations}.
\begin{equation}
\lambda_{\pm}=\pm \cot{\theta}/k(\sqrt{r_p/r_s}+\sqrt{r_s/r_p}),\ |\pm\rangle\sim \begin{pmatrix}
    \sqrt{r_p}\\ \pm i \sqrt{r_s}
\end{pmatrix}
\label{eqeigen}
\end{equation}
Note that the real and imaginary eigenvalues manifest as spatial and angular (momentum-domain) beam shifts, respectively \cite{toppel2013goos,gotte2014eigenpolarizations}. Now we examine the behaviour of $\hat{A}_y$, $\lambda_{\pm}$, and $|\pm\rangle$ with changing system parameters $r_p$, $r_s$, and $\theta$. 
\par
In the case of TIR ($\theta>$ the critical angle $\theta_c$), the Fresnel reflection coefficients consist of only phase factors $r_p^{TIR}=e^{i\delta_p},\ r_s^{TIR}=e^{i\delta_s}$, with $\ \delta_p-\delta_s=\delta$ \cite{born2013principles}. In this case, $\hat{A}_y^{TIR}$ is Hermitian and takes the form $-\cot{\theta}/k\bigg((1+\cos{\delta})\hat{\sigma}_y+\sin{\delta}\hat{\sigma}_x\bigg)$ where $\hat{\sigma}_i,\ i=x,y,z$ are the standard Pauli matrices \cite{shankar2012principles,jayaswal2014observation}. The eigenvalues are real $\lambda_\pm^{TIR}=\pm 2\frac{\cot{\theta}}{k} \cos{(\delta/2)}$ indicating a spatial shift (Fig. \ref{fig1}(a), (b)), and corresponding eigenstates are elliptic and orthogonal $|{\pm}\rangle^{TIR} \sim [e^{i\delta/2}\ \pm i]^T$.

\par
However, peculiarity arises for PR as $r_p^{PR}$ and $r_s^{PR}$ are real, and the corresponding shift matrix $\hat{A}_y^{PR}$ is non-Hermitian.
\begin{equation}
    \hat{A}_y^{PR}=\frac{\cot{\theta}}{k}\bigg(-(1+ \frac{r_p/r_s+r_s/r_p}{2})\hat{\sigma}_y+ i\frac{r_p/r_s-r_s/r_p}{2}\hat{\sigma}_x\bigg)
    \label{eqPRmat}
\end{equation}
Interestingly, \eqref{eqeigen} suggests that the eigenvalues and eigensates of $\hat{A}_y^{PR}$ undergo a transition around $\theta=$ Brewster's angle $\theta_B$ (corresponding to $r_p^{PR}=0$) \cite{gupta2015wave, born2013principles}. Although $\hat{A}_y^{PR}$ is non-Hermitian, at an angle of incidence $\theta>\theta_B$ ($r_p^{PR}<0$), the eigenvalues are still real; however, the corresponding eigenstates are non-orthogonal elliptical \cite{gotte2014eigenpolarizations}. On the other hand, at $\theta<\theta_B$ ($r_p^{PR}>0$), the eigenvalues become imaginary manifested as a momentum domain beam shift (SI Sec. S.6), and the non-orthogonal eigenstates become linear (Fig.\ref{fig1}(a), (b)). This change from imaginary to real eigenvalues essentially resembles a typical PT-transition in a non-Hermitian system around an ``exceptional point'' \cite{gotte2014eigenpolarizations,bender1998real,el2018non,krasnok2021parity}. This was partly alluded to by G\"otte and colleagues \cite{gotte2014eigenpolarizations}. Therefore, the demonstaration of such transformations in the eigenspectrum of the IF shift matrix in PR mimicking a PT-transition appear to be the first outcome of the present study (Fig.\ref{fig1}(a), (b)). More importantly, this work reveals the origin of all these transformations, starting from TIR to PR, as we discuss in Sec.\ref{sec4}

\begin{figure*}[ht!]
\centering
\includegraphics[width=0.8\linewidth]{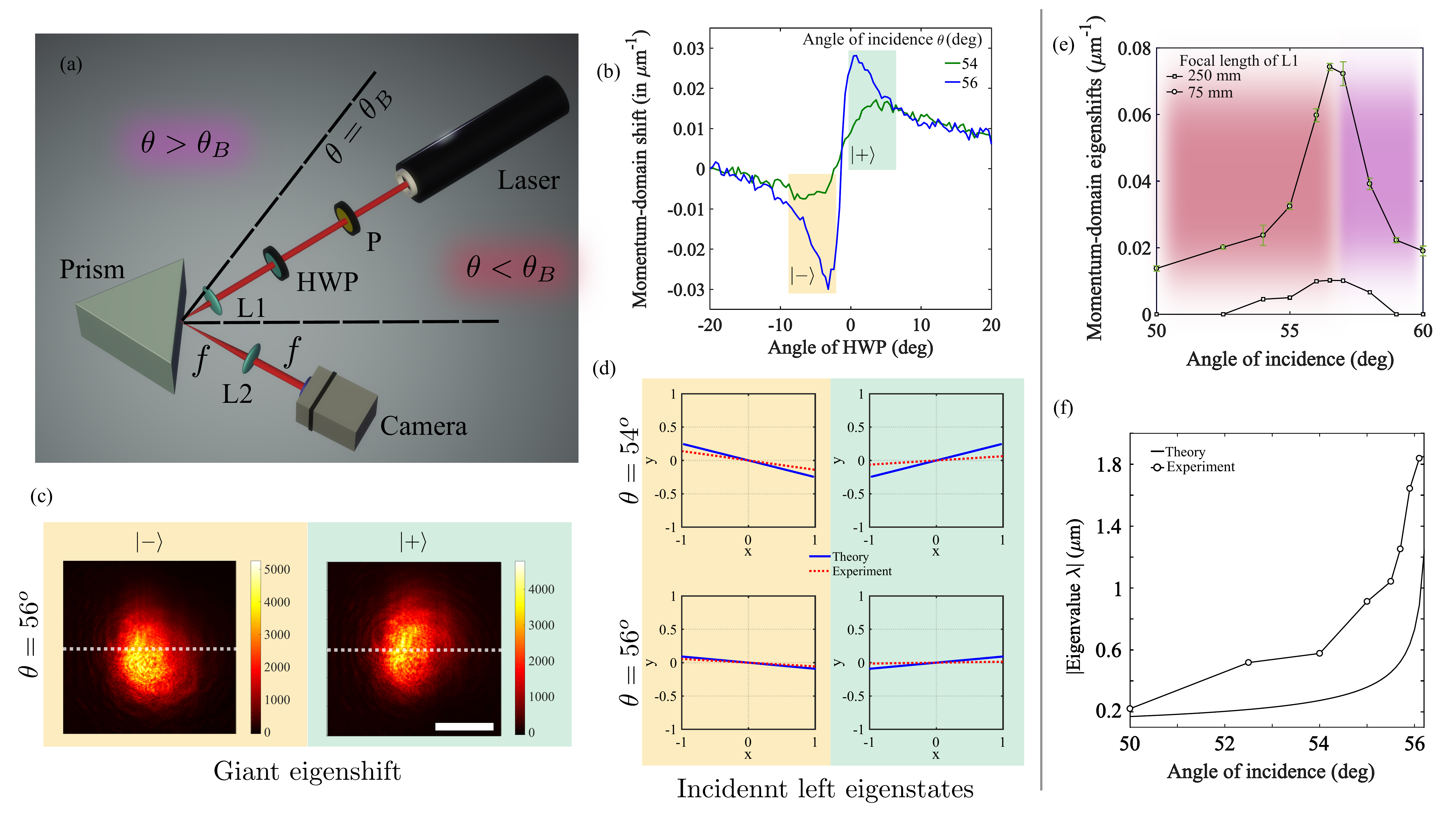}
\caption{\textit{Experimental detection of momentum-domain giant eigenshifts at $\theta < \theta_B$.} (a) Experimental setup. The laser beam from a He-Ne Laser source ($633 nm$) is partially reflected from a glass prism of refractive index $\sim 1.5$. The focal length of the first lens, L1, is used to tune the spot size of the incident beam to get a further control over the magnitude of the momentum domain eigenshift. The second lens, L2 (focal length $100mm$), is used as a Fourier lens to implement a conventional $2-f$ momentum space imaging system \cite{goodman2005introduction}. (b) Variation of the experimentally observed momentum-domain transverse shift with varying angle of the polarization of the input beam for two representative $\theta= 54^{o}$ (green solid line), $56^{o}$ (blue solid line). When the input polarization matches with the left eigenstates $\ket{-}$ (yellow shaded region) and $\ket{+}$ (green shaded region) of $A_y^{PR}$, we get giant shifts, as presented in (c) for $\theta=56^{o}$. Colour bars are intensities in arbitrary units. White scalebar represents a momentum-domain length of $\sim 0.24\mu m^{-1}$. (d) The theoretically calculated left eigenstates match well with experimentally determined (from the orientation of P and HWP) input polarization state. The states are represented in standard $x-y$ Cartesian space. (e) Variation of the momentum-domain eigenshifts with changing angle of incidence for two different focal lengths of L1: $250mm$ (black squares), $75mm$ (black circles). Error bars indicate statistical errors. The colour code follows that of Fig. \ref{fig1}. (f) The variation of retrieved eigenvalues (from a larger dataset of eigenshifts) with changing angle of incidence for $75mm$ focal length of L1, are in agreement with theoretical prediction (\eqref{eqeigen}). The slight mismatch occurs due to error-prone estimation of the beam waist at focus (see Materials and methods).}
\label{fig3}
\end{figure*}
\paragraph{Experimental detection of momentum-domain IF shift for partial reflection}
\label{sec3}
Although the spatial shift at $\theta >\theta_B$ has been detected in \cite{gotte2014eigenpolarizations}, the momentum-domain shifts at $\theta < \theta_B $ are still unexplored. We first experimentally detect these shifts. Note that the typical magnitudes of the eigenvalues of $\hat{A}_y^{PR}$ are small. However, the eigenvalues can directly be measured around singular points, e.g., Brewster's angle, where the magnitude of the eigenvalues becomes large \cite{gotte2014eigenpolarizations,modak2023longitudinal}. A glass prism is used to make a fundamental Gaussian beam partially reflected (Fig.\ref{fig3}(a)). To generate the desired linear eigenstates $\ket{\pm}$ in the incident beam, we put a polarizer (P)-Half wave plate (HWP) combination (Fig.\ref{fig3}(a)), two different focal lengths of L1: $250mm$, and $75mm$ are used (see Materials and methods). When the input polarization state, by tuning P and HWP, matches with the left eigenstates of $\hat{A}_y^{PR}$ (see Materials and methods), the momentum-domain shifts become giant (Fig.\ref{fig3}(b), (c)). The difference in the centroid positions in the momentum-domain of the reflected beam for two eigenstates $|\pm\rangle$, i.e., eigenshifts \cite{modak2022tunable} are plotted with changing angle of incidence in Fig.\ref{fig3}(e).  Due to the increasing magnitude of the imaginary eigenvalues, the momentum-domain eigenshift increases around Brewster's angle \cite{gotte2014eigenpolarizations,gotte2013limits}. As also evident from Fig.\ref{fig3}(e), the eigenshifts increase significantly with reducing the focal lengths of L1, which is a signature of momentum-domain beam shifts \cite{bliokh2013goos,goswami2014simultaneous}. Although the magnitude of such eigenshifts falls rapidly at $\theta>\theta_B$, there is still some appreciable amount of momentum domain shift. This experimental ambiguity is discussed in the SI Sec. S.6. It is also evident that the magnitude of the giant eigenshifts increase further with reducing the focal length of L1 supporting the usual characteristic on any momentum-domain beam shift \cite{bliokh2013goos, hosten2008observation}. The experimentally retrieved eigenvalues (see Materials and methods) agree with the theoretical predictions as noted in \eqref{eqeigen} (see Fig.\ref{fig3}(f)).

\par
\paragraph{Origin of all the peculiarities in the Hermiticity of IF shift}
\label{sec4}
The beam shifts have their origin in the corresponding momentum domain polarization modulation of the beam \cite{bliokh2013goos,toppel2013goos}. The transverse IF shift, specifically, originates from the polarization distribution along the transverse component of momentum ($k_y$) in the beam \cite{bliokh2013goos}, and therefore, we consider a polarized 1-D Gaussian beam consisting only $k_y$. The momentum space evolution of such a vector beam after TIR or PR is governed by a momentum domain Jones matrix $\hat{J}(k_y)$. While $\hat{A}_y$ provides the information of these polarization-dependent beam shifts, the corresponding $\hat{J}(k_y)$ consists of all the information of the momentum-domain polarization transformation of the beam exhibiting beam shift. In fact, the shift matrix is conventionally derived from this momentum domain Jones matrix $\hat{J}(k_y)$ \cite{toppel2013goos,gotte2014eigenpolarizations}, i.e.,
\begin{equation}
\begin{pmatrix} r_p & \frac{-k_y \cot{\theta}}{k}(r_p+r_s)\\ \frac{k_y \cot{\theta}}{k}(r_p+r_s) & r_s
\end{pmatrix}\sim e^{-ik_y\hat{A}_y} \begin{pmatrix} r_p & 0\\0&r_s\end{pmatrix}
\label{eq5}
\end{equation}
Here $\begin{pmatrix} r_p & 0\\0&r_s\end{pmatrix}$ is the zeroth order Fresnel reflection matrix \cite{toppel2013goos,gotte2014eigenpolarizations}, and the diffractive corrections of the beam lead to the correction term $\hat{J}_c(k_y)=e^{-ik_y\hat{A}_y}$ responsible for the beam shift. The real and imaginary eigenvalues of $\hat{A}_y$, therefore, result in the spatial and angular shift of the beam respectively \cite{gotte2014eigenpolarizations}.
\par
Note that $\hat{A}_y$ acts as the corresponding differential Jones matrix of $\hat{J}_c(k_y)$ \cite{jones1948new,modak2021generalized}. Although the conventional Jones algebra has been extensively used in polarization optics \cite{gupta2015wave}, its differential formalism is less noticed \cite{jones1948new}. However, differential Jones matrix formalism is quite profound, particularly due to its structural analogy with the Schr\"{o}dinger's formalism of Quantum mechanics \cite{jones1948new,shankar2012principles}. 
\par 
The eigenvalues of $\hat{J}_c(k_y)$ are
\begin{equation}
    \Lambda_{\pm}=e^{-ik_y(\pm\lambda)};\text{with eigenstates} \ket{\pm}
    \label{eq6}
\end{equation}
 It is well-known that the phase and amplitude eigenvalues of a Jones matrix are associated with polarization retardance (unitary) and diattenuation (non-unitary) effect \cite{gupta2015wave}. Moreover, it is apparent from \eqref{eqeigen} and \eqref{eq6} that the $k_y$-dependent phase and amplitude eigenvalues of $\hat{J}_c(k_y)$ correspond to the real and imaginary eigenvalues of $\hat{A}_y$. The eigenstates of both $\hat{A}_y$ and $\hat{J}_c(k_y)$ are also the same. Hence, real and imaginary eigenvalues of $\hat{A}_y$ are manifestations of momentum domain polarization retardance (phase eigenvalues of $\hat{J}_c(k_y)$) and diattenuation (amplitude eigenvalues of $\hat{J}_c(k_y)$) effects respectively. This one-to-one correspondence between the eigenspectrums of $\hat{J}_c(k_y)$ and $\hat{A}_y$ allows us to use the polarization property of $\hat{J}_c(k_y)$ to understand the physical origin of all the transformations in the eigenspectrum $\hat{A}_y$.
\par
\begin{figure}[h!]
\centering
\includegraphics[width=\linewidth]{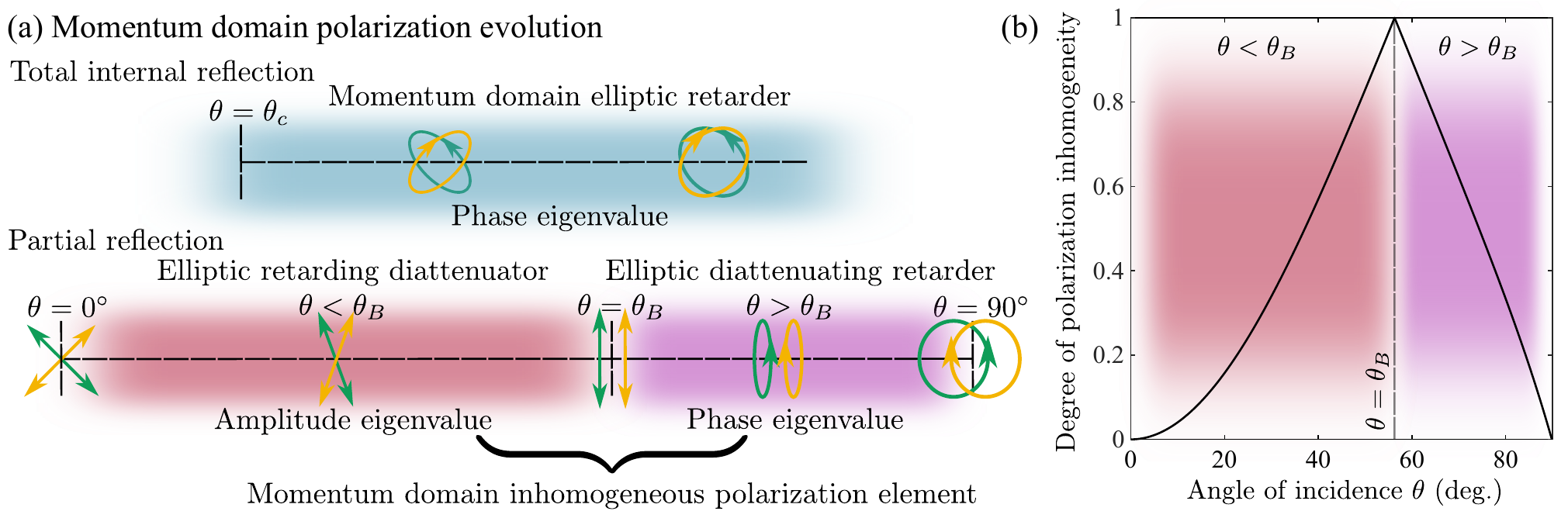}
\caption{\textit{Momentum-domain polarization transformation is the origin of the discussed characteristics of the eigenspectrum (noted in Fig.\ref{fig1}(b)) of transverse optical beam shift}. (a) For TIR, the system acts as a momentum domain elliptic retarder. In PR, the system acts as a momentum-domain inhomogeneous polarization element, transforming from diattenuating retarder to retarding diattenuator around $\theta=\theta_B$. The eigenstates of all the mentioned momentum-domain polarization elements are same as the corresponding beam shift operator (see Fig.\ref{fig1}(b) and \eqref{eq6}). (b) Variation of the degree of polarization inhomogeneity for partial reflection of a beam from an air-glass (refractive index $=1.5$) interface. Brewster's angle ($\theta=\theta_B=56.31^{\circ}$) corresponds to the maximum degree of inhomogeneity. The colour code follows that of Fig. \ref{fig1}.}
\label{fig2}
\end{figure}

For TIR, the Jones matrix $\hat{J}_c^{TIR}(k_y)$ indicates a simultaneous presence of two unitary systems -- circular and $\pm45^{\circ}$ linear retarder \cite{jones1948new,modak2021generalized} in the momentum-domain of the beam (Fig. \ref{fig2}(a)). $\hat{J}_c^{TIR}(k_y)$ (see SI Sec. S.2) is, therefore, unitary and has phase eigenvalues $\Lambda_{\pm}^{TIR}=e^{-ik_y(\pm\lambda^{TIR})}$ with orthogonal elliptic eigenvectors $|\pm\rangle^{TIR}$. The corresponding shift matrix, $\hat{A}_y^{TIR}$ is, therefore, Hermitian having real eigenvalues $\pm\lambda^{TIR}$.
\par
On the contrary, the shift matrix for partial reflection $\hat{A}_y^{PR}$ is non-Hermitian, and can be written as an imaginary combination of $\hat{\sigma}_y$ and $\hat{\sigma}_x$, as mentioned earlier. $i\hat{\sigma}_x$ is the differential matrix for a $\pm45^{\circ}$ linear diattenuator (non-unitary) \cite{jones1948new,modak2021generalized}. Such a simultaneous presence of circular/linear retarder and linear/circular diattenuater mimics an inhomogeneous polarization anisotropy element \cite{lu1994homogeneous} in the momentum ($k_y$) domain. Note that non-orthogonal eigenstates (like $|\pm\rangle$) are one of the signatures of inhomogeneous polarization elements \cite{lu1994homogeneous}. Now we discuss how an inhomogeneous polarization element with the differential matrix $a\hat{\sigma}_{y} + ib\hat{\sigma}_{x}$ ($a,b$ are real) behaves under different conditions on the realtive magnitude of $a$ and $b$. 
\begin{enumerate}[wide, labelwidth=!, labelindent=0pt, noitemsep] 
    \item It acts as a diattenuating retarder at $a^2>b^2$ (SI Sec. S.2), where the retardance effect dominates. Even though the differential matrix is non-Hermitian and the corresponding exponential Jones matrix is non-Unitary \cite{jones1948new,modak2021generalized}, it still has phase eigenvalues. The corresponding eigenstates are non-orthogonal elliptic (Fig.\ref{fig2}(a)). 
    \item At $a^2<b^2$, however, the Jones matrix has amplitude eigenvalues, and eigenstates become non-orthogonal linear (SI Sec. S.2). Amplitude eigenvalues indicate that the system converts into an retarding diattenuater (Fig.\ref{fig2}(a)) \cite{jones1948new,modak2021generalized}. In this case, the diattenuation dominates.
    \item At $a^2=b^2$, the diattenuation and retardance effects contribute equally and give rise to the maximally inhomogeneous polarization element where the eigenstates of the corresponding Jones matrix become collinear (Fig.\ref{fig2}(a)).
\end{enumerate}
Additionally, we observe that such a transition in the eigenspectrum of the Jones matrix is general for an inhomogeneous polarization element with simultaneous retardance and diattenuation (moreover, for an imaginary combination of Pauli matrices; SI Sec. S.4).
\par
As mentioned above, the IF shift in partial reflection is a natural manifestation of such an inhomogeneous polarization anisotropy element in the momentum domain. At an angle of incidence $\theta>$ $\theta_B$, $\hat{J}_c^{PR}(k_y)$ (SI Sec. S.2) has phase eigenvalues resulting in real eigenvalues of $\hat{A}_y^{PR}$ with non-orthogonal elliptic eigenstates $|\pm\rangle^{PR}$ (\eqref{eq6}). Therefore the spatial IF shift in partial reflection for $\theta>\theta_B$ (equivalent to Case 1 above) manifests a momentum domain elliptic diattenuating retarder (Fig.\ref{fig2}(a)). For $\theta<\theta_B$, $\hat{J}_c^{PR}(k_y)$ has amplitude eigenvalues resulting in imaginary eigenvalues of $\hat{A}_y^{PR}$ with non-orthogonal linear eigenstates (\eqref{eq6}, SI Sec. S.2). Therefore the spatial IF shift in partial reflection for $\theta<\theta_B$ (Case 2) is the manifestation of the momentum domain elliptic retarding diattenuator. At $\theta=\theta_B$ (Case 3), the eigenstates become collinear, manifesting the maximum degree of inhomogeneity \cite{lu1994homogeneous}. The variation of the degree of inhomogeneity $1-|r_p^{PR}/r_s^{PR}|$ (SI Sec. S.3) with changing the angle of incidence is plotted in Fig.\ref{fig2}(b). The momentum domain polarization transformation becomes maximally inhomogeneous when $\theta\rightarrow \theta_B$ and acts as a homogeneous polarization element when $\theta\rightarrow 0^{\circ},\ 90^{\circ}$. Therefore, it can be concluded that such transition in the eigenspectrum of the transverse IF shift matrix is a manifestation of the corresponding momentum domain inhomogeneous polarization transformation.
\par
Note that all such polarization-dependent optical beam shifts are essentially rooted in the momentum or space-domain polarization transformation of the beam \cite{bliokh2013goos,toppel2013goos,kong2019goos}. We considered the IF shift in one of the simplest light-matter interactions, i.e., partial reflection of a fundamental Gaussian beam. However, all the abovementioned discussions can be trivially extrapolated for all other complex situations.
\paragraph{PT-unbroken and PT-broken phase of IF shift}
\label{sec5}
The Jones matrix $\hat{J}_c^{PR}(k_y)=e^{-ik_y\hat{A}_y^{PR}}$ denotes a $k_y$ evolution of the input polarization state with $\hat{A}_y^{PR}$ being the corresponding evolution Hamiltonian. Such a non-Hermitian Hamiltonian ($a\hat{\sigma}_y+ib \hat{\sigma}_x$) in the form of a complex combination of different Pauli matrices are, in general, PT-symmetric \cite{gotte2014eigenpolarizations,rath2020generating,wang2010symmetry} (SI Sec. S.5). At $a^2>b^2$, this Hamiltonian is in the PT-symmetric phase and has real eigenvalues \cite{bender1998real,el2018non,krasnok2021parity,ozdemir2019parity}. This situation corresponds to $\theta>\theta_B$ in case of $\hat{A}_y^{PR}$. On the other hand, $a^2<b^2$ is the corresponding PT-broken phase where eigenvalues are imaginary \cite{bender1998real,el2018non,krasnok2021parity,ozdemir2019parity}. In case of $\hat{A}_y^{PR}$, this situation arises at $\theta<\theta_B$. Therefore, the transition from PT-symmetric to PT-broken phase appears at $a^2=b^2$ corresponding to $\theta\rightarrow\theta_B$ for $\hat{A}_y^{PR}$. A general diattenuator-retarder inhomogeneous polarization element and transverse IF shifts both exhibit such PT-transition, as both the systems are represented by similar imaginary combinations of Pauli matrices (SI Sec. S.4, S.5). We further note that the shift matrix $\hat{A}_y^{PR}$ (\eqref{eqPRmat}), in circular basis, i.e., in the eigenbasis of $\hat{\sigma}_y$ holds a similar form to a conventional PT-symmetric Hamiltonian of photonic modes \cite{bender1998real,el2018non,krasnok2021parity,ozdemir2019parity}. 
\begin{equation}
    \hat{A}_y^{PR}= a \hat{\sigma}_z + i b \hat{\sigma}_y
\label{eqcirc}
\end{equation}
As apparent from the structure of $\hat{A}_y^{PR}$ in \eqref{eqcirc}, the equal and opposite polarities of pristine spin-Hall shift $\pm a$ with left and right circular polarization eigenstates mimic the two coupled systems. $b$, comprising linear polarization-dependent losses (the unequal magnitude of $r_p^{PR}$ and $r_s^{PR}$ during partial reflection) in the momentum domain, provides the coupling between the two spin-Hall modes. This hybridization leads to new sets of eigenvectors, i.e., generally elliptical polarization states, which may not be orthogonal. Thus, the parameter $b$ can be identified as the coupling coefficient in the usual scenario of resonant systems having gain and loss \cite{bender1998real,el2018non,krasnok2021parity,ozdemir2019parity}. At $r_p^{PR}=r_s^{PR}$ ($\theta\rightarrow 90^{\circ}$), the shifts become purely spin-Hall analogous to an uncoupled resonant system with real eigenvalue. This is also consistent with the physical origin of the spin-Hall shift \cite{bliokh2013goos}. The spin-Hall shift originates from the equal and opposite magnitude of the momentum domain geometric phase gradient for input left and right circular polarization \cite{bliokh2013goos}. Linear polarization-dependent reflectivities, i.e., $r_p^{PR}$ and $r_s^{PR}$, generate an asymmetry of geometric phase gradient between circular polarization states \cite{bliokh2013goos}. It is apparent that when the coupling is small ($a^2>b^2$), although the eigenstates become non-orthogonal, the pristine spin-Hall system dominates and gives real eigenvalues. However, at $a^2<b^2$, the coupling dominates providing imaginary eigenvalues. Although the transition from real to imaginary eigenvalues happens around Brewster's angle ($a^2=b^2$) at the same time, one can not go arbitrarily close to such a singular point, i.e., $r_p^{PR}=0$ (in our system) as the exponentiation of $\hat{A}_y^{PR}$ can not be done there (\eqref{eq5}). However, it should be noted that in practice, the magnitude of the beam shift at $\theta=\theta_B$ takes finite value \cite{gotte2013limits}, as also apparent from Fig.\ref{fig3}(e). Therefore, the richer physics of the non-Hermitian systems is built in a simple partial reflection of a Gaussian beam.

\par
\paragraph{Discussions: non-Hermitian spin-orbit photonics}
\label{sec5}
 We demonstrate that the observed PT-transition and consequently the existence of real eigenvalues in non-Hermitian IF shift originates from the momentum domain inhomogeneous polarization transformation of the beam.  We characterize the entire parameter space of the IF shift by demarcating the Hermitian, PT-unbroken and PT-broken regimes. We probe the previously unexplored PT-broken regime of IF shift by experimentally detecting the momentum-domain giant eigenshifts. More importantly, the description  involving can momentum domain inhomogeneous polarization transformation can be generalised to all such polarization-dependent beam shifts occurring in different light-matter interactions. Besides resolving this fundamental issue of non-Hermitan optical beam shifts, the present study leaves with some important consequences, as mentioned below.

 \par
The IF shift matrix and the differential matrix of inhomogeneous polarization elements are both modeled through similar imaginary combinations of Pauli matrices that incorporate non-Hermitian physics and exhibit PT-transition. In the context of the former, the momentum-domain polarization modulation, described by $\hat{J}_c(k_y)$, is generally related to the geometric phase evolution and spin-orbit interaction of light already demonstrated in a wide variety of optical systems and metamaterials \cite{bliokh2013goos,bliokh2016spin,bliokh2015spin,cardano2015spin,zhang2023spin}. However, in dealing with such spin-orbit effects, the polarization-dependent losses are usually ignored. Yet in practice, such losses are inherent, which make the system non-Hermitian \cite{bliokh2013goos,bliokh2016spin,bliokh2015spin,cardano2015spin,zhang2023spin}. Our study provides a framework for designing controllable non-Hermitian spin-orbit optical materials by including and, moreover, tailoring such polarization-dependent losses. The latter indicates that PT-symmetric non-Hermitian systems can be constructed using regular polarization optical elements (anisotropic materials) by suitably mixing the diattenuation (polarization-dependent losses) and retardance (polarization-dependent phase modifications) effects \cite{lu1994homogeneous,azzam2016stokes}. Hence, this work presents substantial fundamental advancement in understanding optical beam shift, opens up new directions in the study of non-Hermitian systems, and provides simple platforms of optical beam shifts to realize and investigate inherent physics.

\section*{Acknowledgements} 
The authors thank the support of the Indian Institute of Science Education and Research Kolkata (IISER-K), Ministry of Education, Government of India. The authors would like to acknowledge the Science and Engineering Research Board, Government of India, for the funding (grant No. CRG/2019/005558). We want to acknowledge Subhasish Dutta Gupta (TIFR Hyderabad), Alok Kumar Pan (IIT Hyderabad), and Abir Mondal (IISER-K) for scientific discussions that helped to improve our work. RD acknowledges Ministry of Education, India for PMRF research fellowship grant. SG additionally acknowledges CSIR, Government of India, for research fellowships.

\section*{Methods} \paragraph*{Experimental detection of momentum-domain Eigenshifts} 
We experimentally detect the momentum domain eigenshifts corresponding to the imaginary eigenvalue of $A^{PR}_y$. We put a Glan-Thompson polarizer P (GTH10M-A, ThorLabs, USA), Half waveplate HWP (WPH10M-633, Thorlabs, USA) combination (Fig.\ref{fig3}(a)) fixed at a motorized precision rotation mount (KPRM1E/M, Thorlabs, USA) to generate the desired eigenstates in the incident light. The glass prism is mounted on a precision rotation base (HDR50/M, ThorLabs, USA)  to precisely control the angle of incidence. Since $\hat{A}_y$ is a non-Hermitian matrix and, hence, one can distinguish between its right and left eigenstates. The right and the left eigenstates are $\begin{pmatrix}
    \sqrt{r_p}\\ \pm i \sqrt{r_s}
\end{pmatrix}$ and 
$\begin{pmatrix}
    \sqrt{r_s}\\ \pm i \sqrt{r_p}
\end{pmatrix}$ respectively. To get the right eigenstate as input polarization, we have to prepare the left eigenstate using P and HWP combinations \cite{gotte2014eigenpolarizations,modak2022tunable}. For $250mm$ focal length (LB1056, Thorlabs, USA), we get a bigger beam waist at focus; for $75mm$ focal length (LB1901, Thorlabs, USA), we get a smaller spot size \cite{bliokh2013goos}. It is evident from the previous literature \cite{milonni2010laser} that the bigger the beam waist of the incident beam, the smaller the momentum domain (angular) shift. It is also evident from our experimental results (Fig.\ref{fig3}(e)) that the eigenshifts increase significantly with reducing the focal length of L1.

\paragraph*{Extraction of eigenvalues from the momentum-domain eigenshifts}
The momentum-domain eigenshift $\Delta$ and the magnitude of the eigenvalue $\lambda$ (see \eqref{eqeigen}) are connected through the following relation.
\begin{equation}
    |\lambda|=\frac{\pi w_0^2}{2\Lambda f}\Delta
    \label{eq7}
\end{equation}
Here $w_0$ is the beam waist at focus of L1 on the air-prism interface, $\Lambda$ is the wavelength of the incident light, and $f$ is the focal length of the Fourier lens, i.e., $f=100mm$. \eqref{eq7} is used to retrieve the eigenvalues from experimentally detected eigenshifts $\Delta$ in Fig.\ref{fig3}(f). Error-prone estimation of $w_0$ (quadratic presence in \eqref{eq7}) is the cause of the slight mismatch between experimentally retrieved eigenvalues and corresponding theoretical prediction in \eqref{eqeigen}. 
\par

\bibliography{main}

\end{document}


\title{Supporting Information: Inhomogeneous Polarization Transformation Reveals PT-Transition in non-Hermitian Optical Beam Shift}
\author{Niladri Modak, Swain Ashutosh, Shyamal Guchhait, Sayan Ghosh, Ritwik Dhara, \\Jeeban Kumar Nayak, Sourin Das, Nirmalya Ghosh}
\maketitle

\section{Non-Hermitian Longitudinal Goos-H\"anchen Optical Beam Shift Operator}
The longitudinal optical beam shift operator $\begin{pmatrix}
    -i\frac{\partial \ln r_{p}}{\partial \theta_i} & 0 \\
0 &  -i\frac{\partial \ln r_{s}}{\partial \theta_i}
\end{pmatrix}$ is Hermitian for total internal reflection (TIR) and non-Hermitian for partial reflection (PR). Similar to Imbert-Fedorov (IF) shift matrix, this non-Hermitian nature can also be understood through the corresponding momentum domain polarization transformation of the reflected beam \cite{gotte2014eigenpolarizations,toppel2013goos}. While partially reflecting from an interface, different longitudinal wavevectors ($k_x$) of the incident beam experience different angles of incidence, and therefore, different Fresnel reflection coefficient \cite{bliokh2013goos}. Due to this dispersion of longitudinal wavevectiors, the beam experience a momentum domain polarization transformation \cite{gotte2014eigenpolarizations,toppel2013goos,bliokh2013goos}. In case of TIR, the Fresnel coefficients appears essentially as phase factors \cite{jackson1999classical}, and therefore, the polarization modulation becomes a momentum domain linear retarder. Hence, the corresponding shift matrix is Hermitian with real eigenvalues. On the other hand, for PR, Fresnel coefficients are real and the corresponding polarization modulation becomes a momentum domain linear diattenuator. The shift matrix, in this case, becomes non-Hermitian with imaginary eigenvalues. The eigenstates of the shift matrix, for TIR \cite{jayaswal2014observation} and PR, remains orthogonal horizontal and vertical linear polarizations.

\section{The form of $\hat{J}_c(k_y)$ in total internal reflection and partial reflection}
For TIR,
\begin{equation}
    \hat{J}_c^{TIR}(k_y)= \begin{pmatrix} \cos{(\frac{2k_{y}\cot{\theta}}{k} \cos{\delta/2})} & \frac{(e^{i\delta}+1)}{\sqrt{2(1+\cos{\delta})}}\sin{(\frac{2k_{y}\cot{\theta}}{k} \cos{\delta/2})}\\ -\frac{(e^{i\delta}+1)}{\sqrt{2(1+\cos{\delta})}}\sin{(\frac{2k_{y}\cot{\theta}}{k} \cos{\delta/2})} & \cos{(\frac{2k_{y}\cot{\theta}}{k} \cos{\delta/2})}\end{pmatrix} 
    \label{eqs1}
\end{equation}
All the parameters in Eq. \eqref{eqs1} are defined in the main text. Straight forward calculation shows that the eigenvalues and eigenstates of $\hat{J}_c^{TIR}(k_y)$, respectively, are $\Lambda_\pm^{TIR}=e^{\mp 2i\frac{k_y\cot{\theta}}{k} \cos{(\delta/2)}},\ |{\pm}\rangle^{TIR} \sim [e^{i\delta/2}\ \pm i]^T$.

For PR, at $\theta>\theta_B$,
\begin{equation}
      \hat{J}_c^{PR}(k_y) =
    \begin{pmatrix}
      \cos{\left[\frac{k_y\cot{\theta}}{k}\left(\sqrt{|\frac{r_p}{r_s}|}+\sqrt{|\frac{r_s}{r_p}|}\right)\right]} & \sqrt{|\frac{r_p}{r_s}|}\sin{\left[\frac{\cot{k_y\theta}}{k}\left(\sqrt{|\frac{r_p}{r_s}|}+\sqrt{|\frac{r_s}{r_p}|}\right)\right]} \\
      -\sqrt{|\frac{r_s}{r_p}|}\sin{\left[\frac{k_y\cot{\theta}}{k}\left(\sqrt{|\frac{r_p}{r_s}|}+\sqrt{|\frac{r_s}{r_p}|}\right)\right]} & \cos{\left[\frac{k_y\cot{\theta}}{k}\left(\sqrt{|\frac{r_p}{r_s}|}+\sqrt{|\frac{r_s}{r_p}|}\right)\right]}  
     \end{pmatrix}
     \label{eq}
\end{equation}
The eigenvalues and eigenstates, in this case, are $\Lambda_{\pm}=e^{\mp ik_y\cot{\theta}/k(\sqrt{|r_p/r_s|}+\sqrt{|r_s/r_p|})},\ |\pm\rangle\sim [\sqrt{|r_p|}\ \pm i \sqrt{|r_s|}]^T$.
\par
For partial reflection, at $\theta<\theta_B$, 
\begin{equation}
     \hat{J}_c^{PR}(k_y) =
    \begin{pmatrix}
      \cosh{\left[\frac{k_y\cot{\theta}}{k}\left(\sqrt{|\frac{r_p}{r_s}|}+\sqrt{|\frac{r_s}{r_p}|}\right)\right]} & -\sqrt{|\frac{r_p}{r_s}|}\sinh{\left[\frac{\cot{k_y\theta}}{k}\left(\sqrt{|\frac{r_p}{r_s}|}+\sqrt{|\frac{r_s}{r_p}|}\right)\right]} \\
      \sqrt{|\frac{r_s}{r_p}|}\sinh{\left[\frac{k_y\cot{\theta}}{k}\left(\sqrt{|\frac{r_p}{r_s}|}+\sqrt{|\frac{r_s}{r_p}|}\right)\right]} & \cosh{\left[\frac{k_y\cot{\theta}}{k}\left(\sqrt{|\frac{r_p}{r_s}|}+\sqrt{|\frac{r_s}{r_p}|}\right)\right]}  
     \end{pmatrix}
     \label{eq}
\end{equation}
The eigenvalues and eigenstates, in this case, are $\Lambda_{\pm}=e^{\pm k_y\cot{\theta}/k(\sqrt{|r_p/r_s|}+\sqrt{|r_s/r_p|})},\ |\pm\rangle\sim [\sqrt{|r_p|}\ \pm \sqrt{|r_s|}]^T$.

\section{Inhomogeneous polarization elements}
Polarization elements are usually characterized by the associated polarization anisotropy effects, e.g. linear and circular diattenuation and birefringence \cite{gupta2015wave}.  Regular ideal polarization elements, such as polarizers or waveplates possess single polarization anisotropy effect, i.e., linear diattenuation or retardance, respectively \cite{gupta2015wave}. However, most of the naturally evolving materials come with the simultaneous presence of multiple polarization anisotropy effects \cite{azzam2016stokes,collett1993polarized,ghosh2011tissue,gupta2015wave,gil2022polarized}. Inhomogeneous polarization elements  are those having (a) simultaneous diattenuation and retardance effects (diattenuator-retarder) or (b) different diattenuation effects (diattenuator-diattenuator) with their axes not parallel or perpendicular \cite{lu1994homogeneous}. In the present study, we limit our discussions only to the diattenuator-retarder inhomogeneous polarization elements only. We consider all possible combinations of such inhomogeneous polarization elements as mentioned below.
\begin{enumerate}[wide, labelwidth=!, labelindent=0pt, noitemsep] 
\item circular retarder and $\pm45^{\circ}$ linear diattenuator
\item circular retarder and horizontal-vertical linear diattenuator
\item horizontal-vertical linear retarder and $\pm45^{\circ}$ linear diattenuator
\item $\pm45^{\circ}$ linear retarder and circular diattenuator
\end{enumerate}
Such inhomogeneous polarization elements are described by the corresponding differential matrices \cite{jones1948new}, respectively, listed in the 1st column of Fig. \ref{fig_inhomotable}. Note that there are two other similar inhomogeneous polarization elements, i.e., simultaneous $\pm45^{\circ}$ linear retarder and horizontal-vertical linear diattenuator, and horizontal-vertical linear retarder and circular diattenuator which are similar to respectively, Case 3 and Case 4 above; therefore their characteristics can be extrapolated from Case 3 and Case 4. The magnitude of retardance is $a$, and diattenuation is $b$ for all the cases. The corresponding exponential matrices \cite{jones1948new,modak2021generalized} are the regular Jones matrices for those inhomogeneous polarization elements.
\par
Unlike homogeneous polarization elements, the eigenstates of the Jones matrices of inhomogeneous polarization elements are non-orthogonal. And the degree of inhomogeneity is conventionally described by the overlap between two eigenstates \cite{lu1994homogeneous}. Maximum inhomogeneity appears when two eigenstates become parallel (at $a^2\rightarrow b^2$), and the overlap becomes unity. In this limit, the magnitude of diattenuation tends to become equal to the magnitude of retardance.
\par
In our case of partial reflection, the shift matrix resembles the differential matrix of an inhomogeneous polarization element consisting of a simultaneous circular retarder and $\pm45^{\circ}$ linear diattenuator. The corresponding momentum domain Jones matrix $\hat{J}_c^{PR}(k_y)$ also has non-orthogonal eigenstates. The overlap of the eigenstates, in our case, is $1-|r_p^{PR}/r_s^{PR}|$ which reaches unity at $\theta\rightarrow\theta_B$ indicating the maximum degree of inhomogeneity.

\section{Transition in the eigenspectrum of a general diattenuator-retarder inhomogeneous polarization element}
\begin{figure}[h!]
\centering
\includegraphics[width=\linewidth]{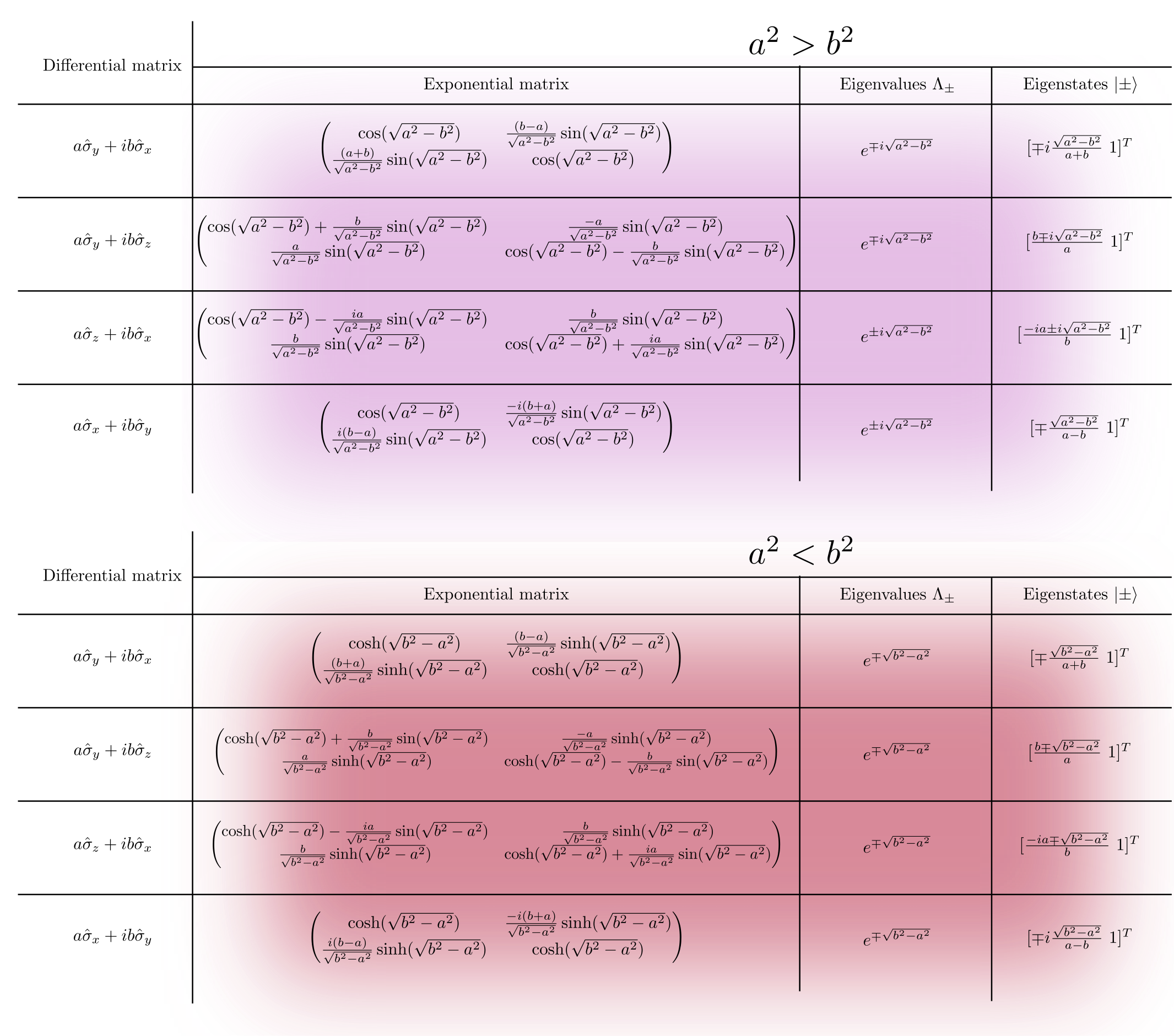}
\caption{\textit{Phase transition in the eigenvalue spectrum of the Jones matrix of general diattenuator-retarder inhomogeneous polarization elements.} The exponential Jones matrices have phase eigenvalues at $a^2>b^2$ and amplitude eigenvalues at $a^2<b^2$ as in these regions retarder and diattenuator, respectively dominate. The eigenstates tend to be collinear at $a^2\rightarrow b^2$.}
\label{fig_inhomotable}
\end{figure}

As discussed in the main text, the shift matrix $A_{y}^{PR}$ is a complex combination of a circular retarder and a $\pm 45^{\circ}$ linear diattenuator. In such an inhomogeneous polarization element, the eigenvalues and eigenstaes of the corresponding Jones matrix exhibit a phase transition around $a^2\rightarrow b^2$ (see Fig. \ref{fig_inhomotable}). In a similar manner, we examine the outcomes of all such possible complex combinations of retarder and diattenuator matrices. The Jones matrices, eigenvalues, and eigenstates are noted for all such combinations in Fig. \ref{fig_inhomotable}.

\section{PT-symmetry in $\hat{A}_y$ for partial reflection}
As discussed in the main text, the transverse optical shift matrix for partial reflection $\hat{A}_y^{PR}$ is non-Hermitian. This non-Hermitian matrix is parity-time (PT) symmetric \cite{gotte2014eigenpolarizations,rath2020generating,wang2010symmetry,ozdemir2019parity}. P operator, in this case, is the third Pauli matrix $\hat{\sigma}_z$. On the other hand, the T operator is just a complex conjugation \cite{bender1998real,krasnok2021parity}. At $\theta>\theta_B$, the eigenvalues of $\hat{A}_y^{PR}$ are real, and corresponding elliptical eigenstates are the eigenstates of the PT operator. Therefore, $\theta>\theta_B$ is the regime for the PT-unbroken phase. On the contrary, at $\theta<\theta_B$, the eigenvalues are imaginary, and corresponding linear eigenstates are not the eigenstates of the PT operator, indicating the PT-broken phase.
\par
\begin{figure}[h!]
\centering
\includegraphics[width=\linewidth]{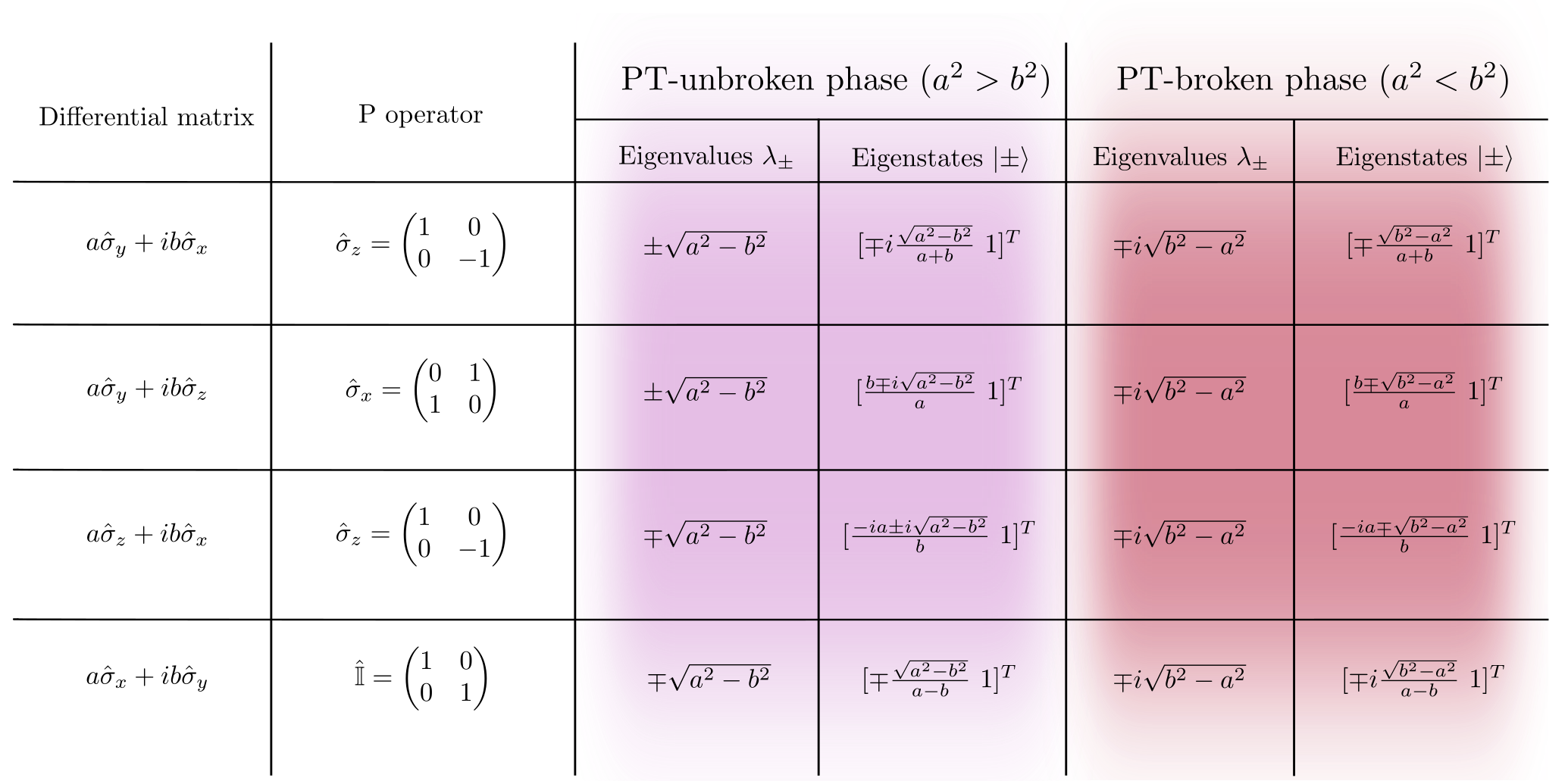}
\caption{\textit{PT-unbroken and broken phase for general diattenuator-retarder inhomogeneous polarization elements.} The forms of the differential matrices of the polarization elements and corresponding parity (P) operators are listed in the first and second columns, respectively \cite{gotte2014eigenpolarizations,rath2020generating,wang2010symmetry,ozdemir2019parity}. Eigenvalues of the differential matrices real at $a^2>b^2$ and corresponding eigenstates are the eigenstates of the PT operator. This regime is the PT-unbroken phase. At $a^2<b^2$, the eigenvalues become imaginary, and the eigenstates are not the eigenstates of the PT operator, indicating the PT-broken phase. The eigenvalues coalesce to zero at $a^2\rightarrow b^2$, and corresponding eigenstates become parallel. Thus, $a^2= b^2$ acts as an exceptional point.}
\label{fig_PTtable}
\end{figure}

Note that $\hat{A}_y^{PR}$ resembles the differential Jones matrix for the circular retarder-$\pm45^{\circ}$ linear diattenuator inhomogeneous polarization element (see Sec. S.3). The differential matrices of all other inhomogeneous polarization elements are also PT-symmetric with the P operators listed in Fig. \ref{fig_PTtable} and the T operator to complex conjugation \cite{el2018non}. In all the cases, eigenvalues of the differential matrices are real at $a^2>b^2$ and corresponding eigenstates are the eigenstates of the PT operator. This regime is the PT-unbroken phase. At $a^2<b^2$, the eigenvalues become imaginary, and the eigenstates are not the eigenstates of the PT operator, indicating the PT-broken phase. The eigenvalues coalesce to zero at $a^2\rightarrow b^2$, and corresponding eigenstates become parallel. Thus, the maximum inhomogeneity situation $a^2= b^2$ acts as an exceptional point (see Sec. S.4).
\par
As mentioned earlier, similar transitions also happen around $\theta\rightarrow\theta_B$ in the case of partial reflection. The maximum degree of inhomogeneity in the corresponding momentum domain Jones matrix $\hat{J}_c^{PR}(k_y)$ (see Fig. 2 of the main text). The eigenstates of $\hat{A}_y^{PR}$ also tend to coalesce; however, the eigenvalues tend to diverge. This happens because the form of $\hat{A}_y^{PR}$ does not allow us to write it as an imaginary combination of $\hat{\sigma}_x$ and $\hat{\sigma}_y$ and moreover, the first order approximation to describe the beam shifts becomes invalid and consequently, the eigenspectrum of $\hat{A}_y^{PR}$ encounters a singularity. 

\clearpage
\section{Experimental ambiguities on the detection of momentum-domain Eigenshifts}
 Due to the increasing magnitude of the eigenvalues, the eigenshift increases around Brewster's angle. Ideally, it should be maximum at $\theta=\theta_B$ and fall to zero just above the $\theta_B$ as the eigenstates are no longer linear in that region (see main text). But in the experimental scenario, the incident beam also has a momentum distribution in the longitudinal direction ($k_x$). Due to this spread, there is always some part of the beam that gets incident at $\theta<\theta_B$ even though the central wavevector of the beam falls at $\theta>\theta_B$. This gives rise to a momentum domain eigenshift even at $\theta>\theta_B$. However, the magnitude of the shift falls more rapidly in this region when one goes off from $\theta=\theta_B$ (see main text Fig. 2(e)). The longitudinal spread reduces significantly for the $250mm$ focal length of L1. Therefore, the magnitude of the angular eigenshift falls more quickly at $\theta>\theta_B$ in this case; however, the magnitudes of the eigenshifts, in this case, are lower.

\bibliographystyle{ieeetr} 
\bibliography{supporting}